\documentclass[prb,aps,epsf,twocolumn,longbibliography]{revtex4-1}
\usepackage{times}
\usepackage{graphicx}
\usepackage{float}
\usepackage{latexsym,amsmath,amssymb,bm,euscript}
\usepackage{color}
\usepackage{epstopdf}
\usepackage[colorlinks=true,linkcolor=blue,citecolor=blue,urlcolor=blue]{hyperref}
\usepackage{hyperref}

\begin{document}

\title{Characterizing Random-singlet State in Two-dimensional Frustrated Quantum Magnets and Implications for the  Double Perovskite Sr$_2$CuTe$_{1-x}$W$_{x}$O$_6$}
\author{Huan-Da Ren$^{1,2}$}
\author{Tian-Yu Xiong$^1$}
\author{Han-Qing Wu$^3$}
\email{wuhanq3@mail.sysu.edu.cn}
\author{D. N. Sheng$^4$}
\email{donna.sheng1@csun.edu}
\author{Shou-Shu Gong$^1$}
\email{shoushu.gong@buaa.edu.cn}
\affiliation{
$^1$Department of Physics, Beihang University, Beijing, 100191, China\\
$^2$Department of Physics, Tsinghua University, Beijing, 100084, China\\
$^3$Guangdong Provincial Key Laboratory of Magnetoelectric Physics and Devices, School of Physics, Sun Yat-Sen University, Guangzhou 510275, China\\
$^4$Department of Physics and Astronomy, California State University Northridge, Northridge, California 91330, USA}

\begin{abstract}
Motivated by experimental observation of the non-magnetic phase in the compounds with frustration and disorder, we study the ground state of the spin-$1/2$ square-lattice Heisenberg model with randomly distributed nearest-neighbor $J_1$ and next-nearest-neighbor $J_2$ couplings.
By using the density matrix renormalization group (DMRG) calculation on cylinder system with circumference up to $10$ lattice sites, we identify a disordered phase between the N\'eel and stripe magnetic phase with growing $J_2 / J_1$ in the presence of strong bond randomness.
The vanished spin-freezing parameter indicates the absence of spin glass order.
The large-scale DMRG results unveil the size-scaling behaviors of the spin-freezing parameter, the power-law decay of the 
average spin correlation, and the exponential decay of the typical spin correlation, which all agree with the corresponding behavior in the one-dimensional random singlet (RS) state and characterize the RS nature of this disordered phase.
The DMRG simulation also provides insights and opportunities for characterizing a class of non-magnetic states in two-dimensional frustrated magnets with disorder.  
We also compare with existing experiments and suggest more measurements for understanding the spin-liquid-like behaviors in the double perovskite Sr$_2$CuTe$_{1-x}$W$_{x}$O$_6$.
\end{abstract}
\maketitle

{\it Introduction.} Spin liquid (SL) is an exotic quantum liquid state realized in frustrated magnets~\cite{lacroix2013, balents2010, savary2016, Norman2016, ZYi2017}, which exhibits long-range entanglement and fractionalized excitations~\cite{wen1991, read1991, chen2010} and may have potential applications in quantum computation~\cite{kitaev2006}.
After extensive search for decades, spin-liquid-like behaviors have been reported in frustrated antiferromagnets~\cite{savary2016, Norman2016, ZYi2017}. 
In particular, most of the SL candidates have been characterized as gapless spin-liquid states in experiment~\cite{Norman2016, fak2018, shimizu2003, yamashita2008, li2015tri, li2015, shen2016, paddison2017}.
Nevertheless, theoretical studies have only established SL states in few highly frustrated models including kagome antiferromagnet and triangular-lattice Heisenberg models with competing interactions~\cite{savary2016, ZYi2017}.
Only considering frustrated interactions may be  insufficient to account for the widely observed spin-liquid-like behaviors in materials.

Another common factor that may suppress magnetic order is disorder, which naturally exists in materials~\cite{shimizu2003, yamashita2008, freedman2010, zhu2014}.
In one dimension (1d), the strong-disorder renormalization group has established the infinite-randomness fixed point (IRFP) with infinite dynamic exponent, which describes the random singlet (RS) state in random spin chains~\cite{ma1979, dasgupta1980, bhatt1982, fisher1992, fisher1994, fisher1995, melin2002, refael2002, shu2016}.
A characteristic feature of the 1d RS state is the drastic difference between the average and the typical spin correlations, which follow the $r^{-2}$ power-law and the exponential decay respectively as a function of distance $r$ due to the logarithmically broad probability distribution of spin correlation~\cite{fisher1994}.
By extensive experimental studies on random spin chain compounds~\cite{bulaevski1972,tu1996,niazi2001,tota2003,shiroka2011,shiroka2019, volkov2020}, recently the RS behaviors have been reported in the compounds BaCu$_2$(Si$_{1-x}$Ge$_x$)$_2$O$_7$ and Ba$_5$CuIr$_3$O$_{12}$~\cite{shiroka2019, volkov2020}.
In two dimensions (2d), it has been shown in experiment that increased disorder can also melt magnetic order and induce spin-liquid-like behaviors~\cite{ono2005, katayama2015, furukawa2015, yamaguchi2017, savary2017}.
However, understanding disorder effect in 2d frustrated systems is a longstanding challenge for theoretical study.
Although the IRFP has been found in few 2d systems~\cite{motrunich2000, kovacs2011, vojta2009pre, vojta2009prb}, the disorder-induced states in general frustrated Heisenberg models are quite elusive~\cite{oitmaa2001, lin2003, kimchi2018}.
Recently, systematic exact diagonalization (ED) studies have unveiled a disordered phase driven by random couplings in various frustrated Heisenberg models, which is conjectured to be  a 2d RS state~\cite{TSakai2014, kawamura2014, Kawamura2015, Kawamura2017, wu2019, uematsu2018, kawamura2019}.
The ED results of the dynamical and thermodynamic properties of this state qualitatively agree with the predictions of gapless spin liquids~\cite{singh2010, TSakai2014, kawamura2014, Kawamura2015, Kawamura2017, wu2019, uematsu2018, kawamura2019}.
Nonetheless, the characteristic properties of the RS state such as the scaling behaviors of spin correlations have not been addressed and it is unresolved whether this disordered phase is indeed a RS state or not.
A recent quantum Monte Carlo (QMC) study on the random $J-Q$ model find a disordered state with the average spin correlation decaying algebraically as $r^{-2}$, which is proposed as the 2d analog of the RS state although the dynamic exponent is finite~\cite{liu2018}.
It is also conjectured that such a state should also exist in the frustrated 2d Heisenberg spin systems with randomness~\cite{liu2018}, which however has not been identified.
Therefore, understanding the spin correlation behaviors is highly desired to identify the nature of the disorder-induced exotic state in 2d frustrated systems.

In this Letter, we use the large-scale density matrix renormalization group (DMRG) calculation to address the characteristic behaviors of spin correlations in the disorder-induced phase.
We study the spin-$1/2$ square-lattice Heisenberg model with the random nearest-neighbor (NN) $J_1$ and next-nearest-neighbor (NNN) $J_2$ couplings.
This system may be considered as the preliminary description of the spin-liquid-like phase in the double perovskite Sr$_2$CuTe$_{1-x}$W$_{x}$O$_6$~\cite{mustonen2018, mustonen2018prb, watanabe2018, vasala2014, babkevich2016, walker2016, koga2016, katukuri2020,hong2020}, which realizes simultaneous tuning of frustration and disorder by the random Te-W cation mixing.
The pure $J_1 - J_2$ square Heisenberg model has a non-magnetic phase near $J_2 / J_1 = 0.5$ due to strong frustration~\cite{chandra1988, sachdev1990, gong2014, liu2022}.
With random couplings, recent ED study has found evidence for a disordered phase and a possible spin glass phase~\cite{uematsu2018}, without resolving the nature of the disordered phase.
Based on the DMRG results, we identify a disordered phase without any magnetic order or spin glass order~\cite{oitmaa2001} in the strong randomness regime.
Furthermore, we unveil the $L^{-1/2}_x$ scaling of the spin-freezing parameter with system length $L_x$, the $r^{-2}$ power-law decay of the average spin correlation, and the exponential decay of the typical spin correlation, which all agree with the corresponding behavior in the 1d RS state~\cite{fisher1994} and characterize the RS nature of this disordered state.
The consistent $r^{-2}$ behavior of average spin correlation also indicates that this state may belong to the same fixed point as the proposed RS state in the $J-Q$ model, which demonstrates the robust universal behavior for such interacting and disorder systems.
The vanished spin-freezing parameter in our model study implies that the spin glass order should be absent in Sr$_2$CuTe$_{1-x}$W$_{x}$O$_6$ at low temperature.
Our results agree with the observations in the muon spin rotation ($\mu$SR) measurements of Sr$_2$CuTe$_{1-x}$W$_{x}$O$_6$, including the absent spin freezing for $x = 0.5$ at very low temperature~\cite{mustonen2018} and the large dynamic exponent supporting the intermediate RS phase~\cite{hong2020}.

\begin{figure}[htp]
\includegraphics[width=0.3\textwidth]{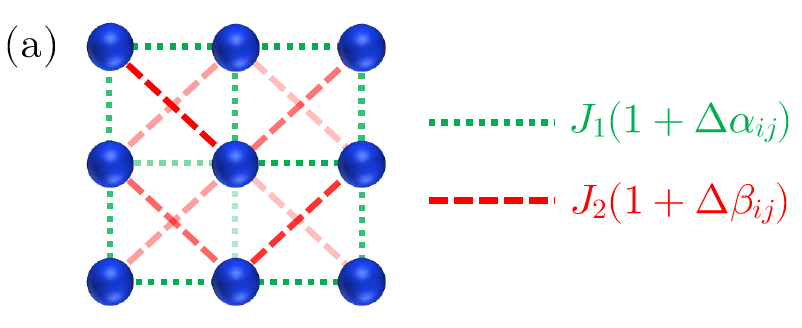}
\includegraphics[width=0.4\textwidth]{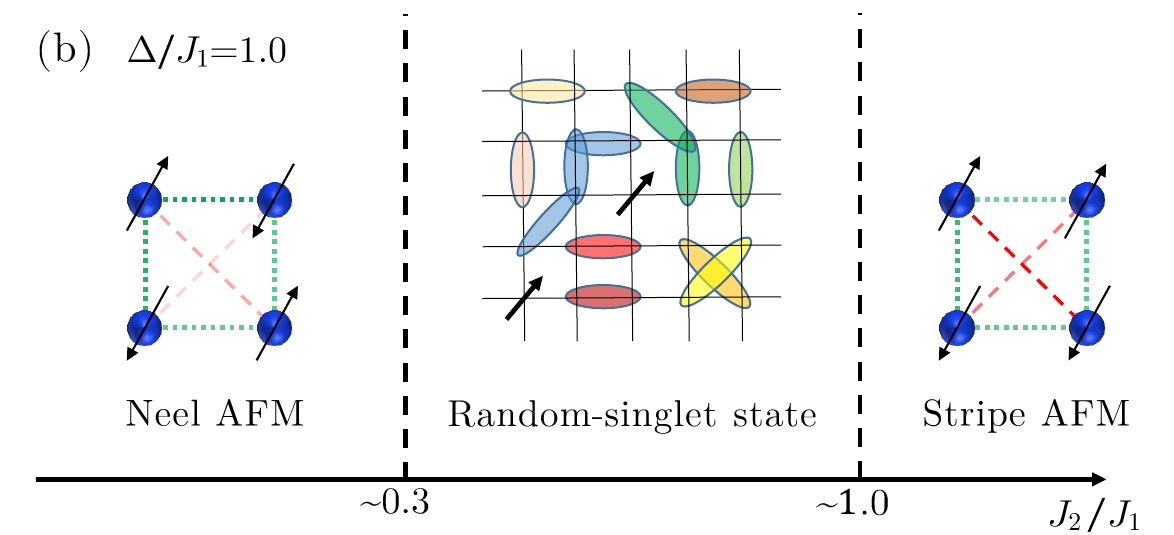}
\caption{Model Hamiltonian and quantum phase diagram of spin-$1/2$ $J_1 - J_2$ square-lattice Heisenberg model with random couplings. (a) The random NN $J_1$ (green bonds) and NNN $J_2$ (red bonds) couplings are uniformly distributed in the interval $\alpha_{ij}, \beta_{ij} \in [-1, 1]$. $\Delta$ is the strength of the randomness. (b) Quantum phase diagram of the model with growing $J_2 / J_1$ and fixed $\Delta / J_1 = 1$, which shows a random-singlet phase between the N\'eel and stripe magnetic phase.}
\label{fig:phase}
\end{figure}

The model with random $J_1, J_2$ couplings is defined as
\begin{equation}
H = \sum_{\langle ij \rangle}J_{1}(1+\Delta \alpha_{ij}){\mathbf{S}}_{i} \cdot {\mathbf{S}}_{j} + \sum_{\langle\langle ij \rangle\rangle}J_{2}(1+\Delta \beta_{ij}) {\mathbf{S}}_{i} \cdot {\mathbf{S}}_{j},
\label{Eq:J1J2model}
\end{equation}
where $\alpha_{ij}$ and $\beta_{ij}$ denote the random variables uniformly distributing in the interval $[-1,1]$, and $\Delta$ controls the randomness strength of the interval $[J_{i}(1-\Delta), J_{i}(1+\Delta)]$ ($i=1, 2$).
We set $J_1 = 1.0$ and choose $\Delta / J_1 = 1.0$ to focus on the strong randomness case, as shown in Fig.~\ref{fig:phase}.
By using the DMRG with spin SU(2) symmetry~\cite{white1992, su2}, we simulate the system on the cylinder geometry with the periodic boundary conditions along the circumference direction ($y$) and the open boundary conditions along the axis direction ($x$), with $L_y$ and $L_x$ being the numbers of sites along the two directions.
We study the systems with $L_y$ up to $10$.
To avoid edge effects, we choose $L_x = 2 L_y$ in most calculations and compute physical quantities using the middle $L_y \times L_y$ subsystem, which is found valid by the agreement of our data with the QMC result. 
We keep $3000$ SU(2) states (equivalent to about $12000$ U(1) states) to ensure the truncation error smaller than $1 \times 10^{-5}$.
We take $100$ ($50$) random samples for $L_y = 4, 6$ ($L_y = 8, 10$), which ensure the good sample average of the quantities; see Supplemental Materials (SM)~\cite{suppl}.
We use ``$\langle \rangle$" and ``[ ]" to represent quantum mechanical and stochastic averages, respectively.

\begin{figure}[htb]
\includegraphics[width=0.239\textwidth]{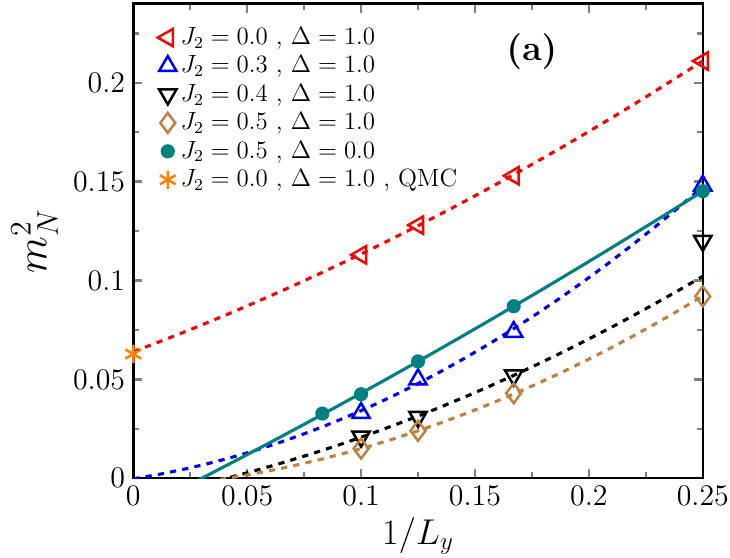}
\includegraphics[width=0.239\textwidth]{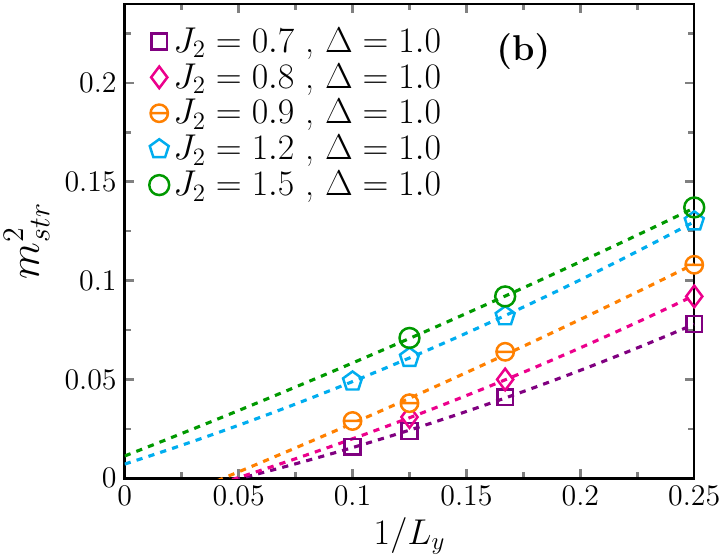}
\includegraphics[width=0.236\textwidth]{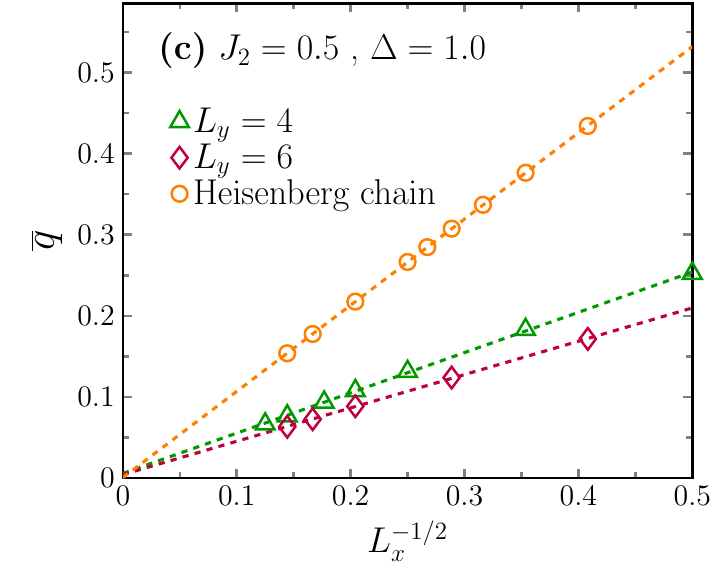}
\includegraphics[width=0.236\textwidth]{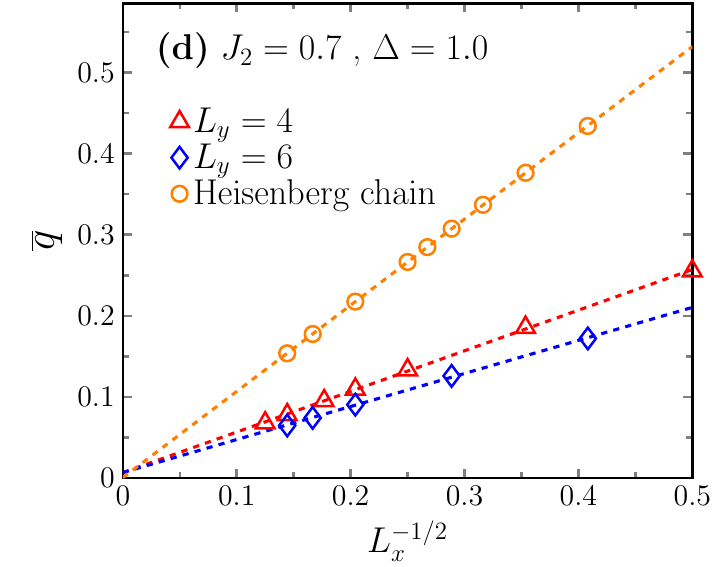}
\includegraphics[width=0.239\textwidth]{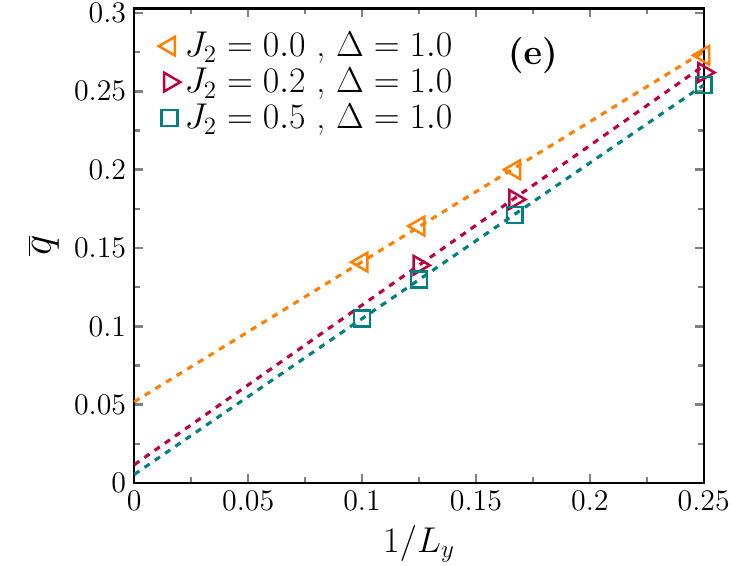}
\includegraphics[width=0.239\textwidth]{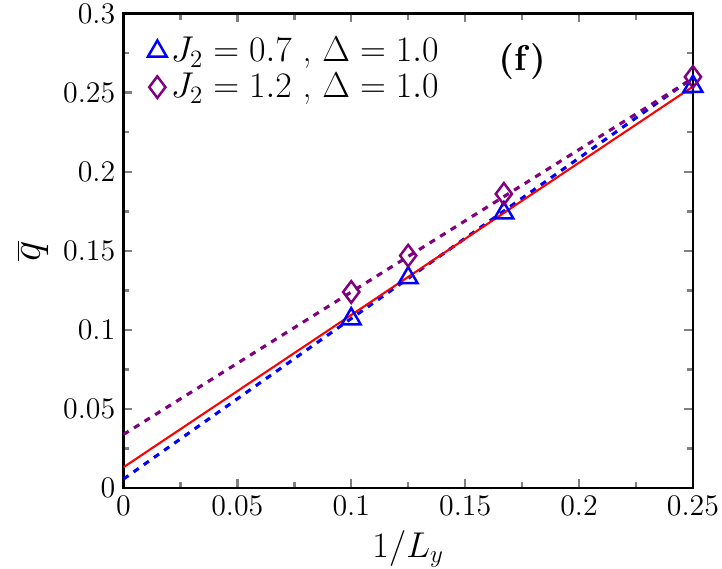}
\caption{Finite-size scaling of order parameters.
(a) and (b) are the N\'eel ($m^2_{\rm N}$) and stripe ($m^2_{\rm str}$) order parameters versus $1/L_y$. For a comparison, we include the result of $J_2 = 0.5, \Delta = 0$~\cite{gong2014}, which has been identified as a non-magnetic state. The lines denote the polynomial fitting of the data up to the second order of $1/L_y$. The star symbol denotes the QMC result~\cite{laf2006}. 
(c) and (d) are the spin-freezing parameter $\overline{q}$ versus $L^{-1/2}_x$ with $L_y = 4, 6$. The result of the 1d RS state of the Heisenberg chain is also shown. The good linear fittings indicate $\overline{q} \propto L^{-1/2}_x$. (e) and (f) are the spin-freezing parameter $\overline{q}$ versus $1 / L_y$ obtained in the middle $L_y \times L_y$ subsystem. The dashed lines denote the linear fitting using the two largest-size data. The solid line in (f) shows the linear scaling of the small-size data with $L_y = 4$ and $6$.}
\label{fig:spin}
\end{figure}

{\it Phase diagram and absent spin glass order.} 
We first determine the phase diagram of the system by computing magnetic order and spin-freezing parameters.
We define magnetic order parameter at the wave vector ${\bf k}$ as
\begin{equation}
m^2({\bf k}) = \frac{1}{N_s^2} \sum_{i, j}\left[ \langle {\bf S}_{i} \cdot {\bf S}_{j} \right \rangle] e^{i {\bf k} \cdot ({\bf r}_{i} - {\bf r}_{j})},
\end{equation}
where $N_s = L_y \times L_y$.
The N\'eel and stripe magnetic order parameters can be defined as $m^2_{\rm N} = m^2(\pi, \pi)$ and $m^2_{\rm str} = [ m^2(0, \pi) + m^2(\pi,0) ] / 2$, where the results at $(0,\pi)$ and $(\pi,0)$ are averaged to reduce geometry effect.
In Fig.~\ref{fig:spin}(a-b), we show the polynomial size scaling of $m^2_{\rm N}$ and $m^2_{\rm str}$.
For $J_2 = 0$, $m^2_{\rm N}$ is smoothly extrapolated to $0.064$ in the $N \rightarrow \infty$ limit, which agrees with the QMC data quite well~\cite{laf2006} and shows the validity of our DMRG setup and calculation.
With growing $J_2$, the N\'eel order seems to be melted at $J_2 \simeq 0.3$, where the N\'eel order parameters are smaller than those of the non-magnetic state at $J_2 = 0.5, \Delta = 0$~\cite{gong2014,liu2022}.
For the larger $J_2$, $m^2_{\rm str}$ becomes finite at $J_2 = 1.2, 1.5$ (see SM for the results of $m^2(0, \pi)$ and $m^2(\pi, 0)$~\cite{suppl}), showing the stripe order survived from randomness.
The strong average spin correlations at $J_2 = 1.2$ also support the emergent stripe order~\cite{suppl}.
Based on the DMRG results, we identify an intermediate non-magnetic phase.

We further explore a possible spin glass phase in this non-magnetic region~\cite{lin2003, uematsu2018}. Spin glass order can be characterized by the spin-freezing parameter $\overline{q}$~\cite{binder1986} defined as
\begin{equation} \label{eq:freezing}
\overline{q} = \frac{1}{N_s} \sqrt{\sum_{i j} \left[ \langle {\bf S}_{i} \cdot {\bf S}_{j} \rangle^2 \right]}.
\end{equation}
If spin orientations freeze, $\overline{q}$ would be nonzero in the thermodynamic limit.
First of all, we show the size scaling of $\overline{q}$ versus $L_x$ for given $L_y$ ($N_s = L_x \times L_y$ in Eq.~\eqref{eq:freezing}) in Fig.~\ref{fig:spin}(c-d).
On the $L_y = 4, 6$ systems with large $L_x$, $\overline{q}$ clearly shows the $L_x^{-1/2}$ scaling behavior, agreeing with that in the 1d RS state.
We further study the decay behavior of $[\langle {\bf S}_i \cdot {\bf S}_j \rangle^2]$ as a function of $|i - j|$, which indeed supports this $L_x^{-1/2}$ scaling behavior~\cite{suppl}.
We also notice that for any given $L_x$, $\overline{q}$ decreases with growing $L_y$, strongly indicating the absent spin freezing in large-size limit.
To confirm this conclusion, we also compute $\overline{q}$ in the middle $L_y \times L_y$ subsystem and analyze $\overline{q}$ versus $1/L_y$ as shown in Fig.~\ref{fig:spin}(e-f).
Since the $\overline{q}$ data decrease slightly faster than linear behavior, we linearly extrapolate the results using the two largest-size data to reduce finite-size effect.
In the magnetic order phases, $\overline{q}$ goes to finite value as expected.
In the non-magnetic region, we can also obtain a very small finite $\overline{q}$ that agrees with the ED conclusion~\cite{uematsu2018} if we use linear scaling for small-size results, as shown by the solid line in Fig.~\ref{fig:spin}(f).
However, by fitting the larger-size data, $\overline{q}$ tends to vanish for $L_y \rightarrow \infty$, which agrees with the results shown in Fig.~\ref{fig:spin}(c-d).
The finite $\overline{q}$ in the ED calculation~\cite{uematsu2018} should be owing to the larger size effect.

\begin{figure}[htb]
\includegraphics[width=0.239\textwidth]{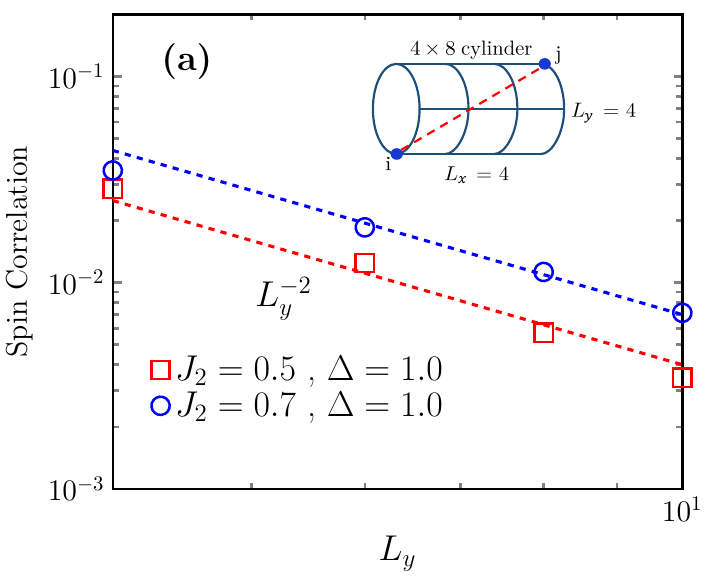}
\includegraphics[width=0.239\textwidth]{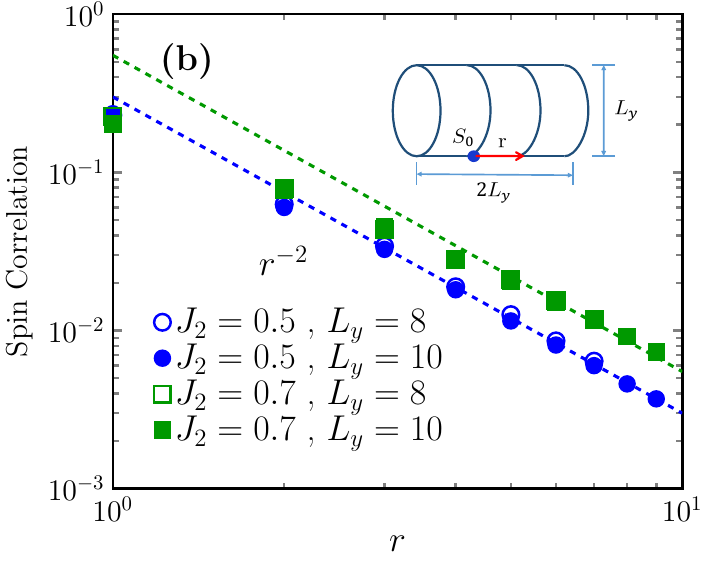}
\includegraphics[width=0.239\textwidth]{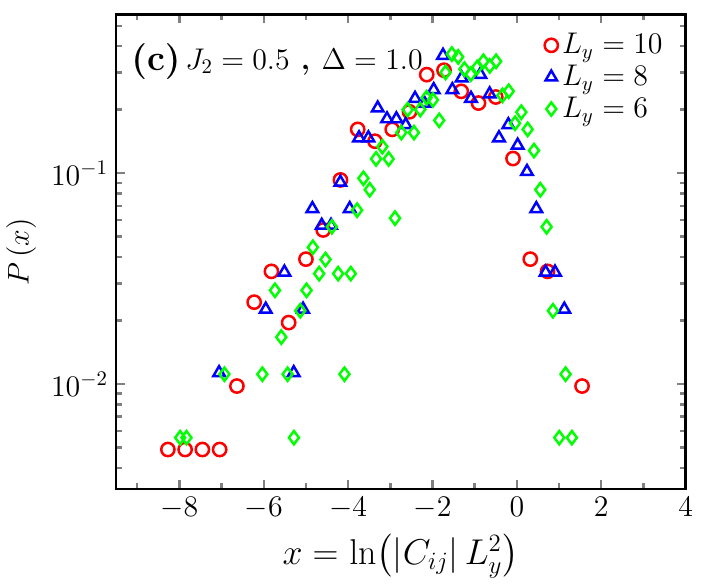}
\includegraphics[width=0.239\textwidth]{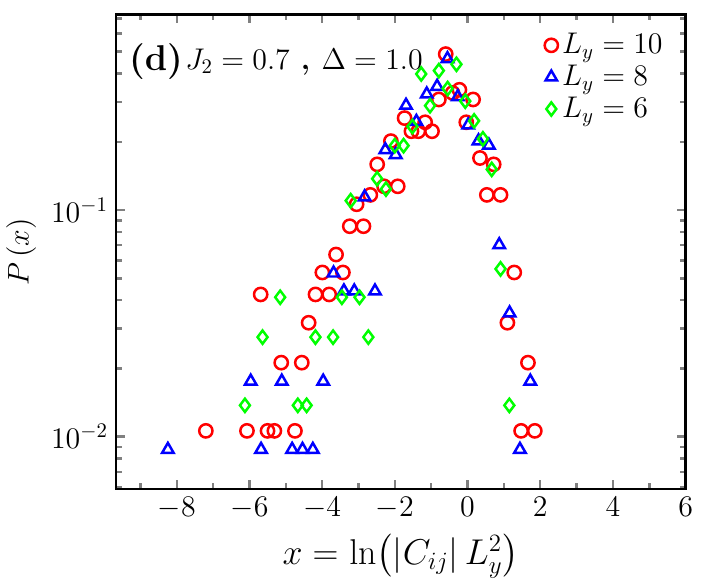}
\includegraphics[width=0.236\textwidth]{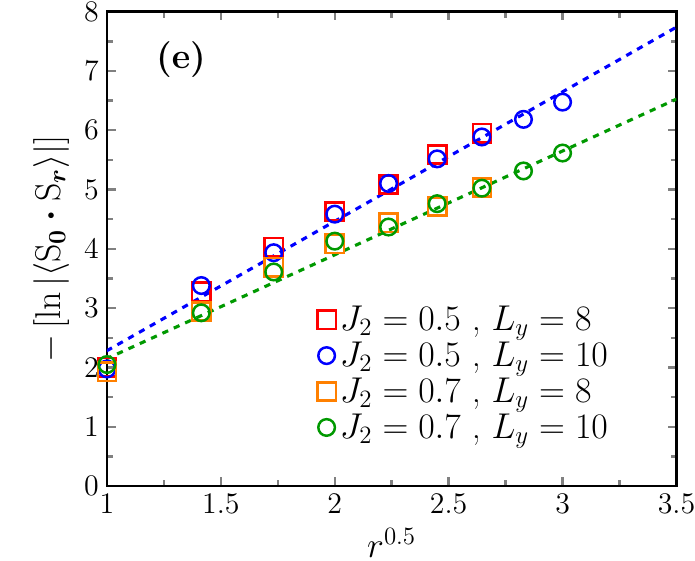}
\includegraphics[width=0.236\textwidth]{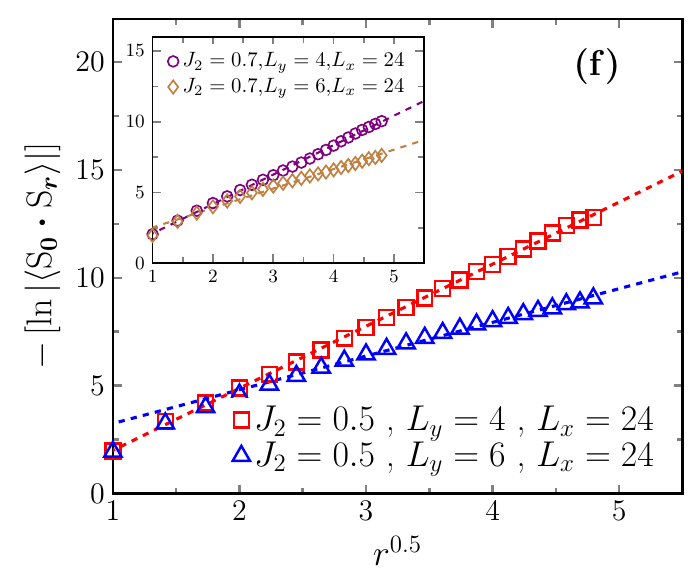}
\caption{Size scaling and distribution of spin correlation functions.
(a) Log-log plot of the average absolute spin correlation versus $L_y$, which are defined for the largest-distance sites on the middle $L_y \times L_y$ subsystem as shown by the example in the inset.
The dashed lines denote the fittings with $C_{s}(L_y) \sim L_y^{-2}$.
(b) Log-log plot of the average absolute spin correlation $\left [ \left | \left \langle {\bf S}_{0}\cdot {\bf S}_{r} \right \rangle \right | \right ]$ versus distance $r$ along the axis direction of cylinder. The reference site ${\bf S}_0$ is defined on the left side of the middle $L_y \times L_y$ subsystem, and the site ${\bf S}_r$ is counted from the left to the right side, as shown by the inset. The dashed lines denote the fittings with $C_{s}(r) \sim r^{-2}$. (c) and (d) show the data collapse of $P(x)$ versus $x = \ln(|C_{ij}| L^{2}_{y})$ on different system sizes.
(e) and (f) show the linear plot of $- [\ln |\langle {\bf S}_0 \cdot {\bf S}_r \rangle|]$ versus $r^{1/2}$.
} 
\label{fig:cor}
\end{figure}

{\it Average and typical spin correlations.} 
Next, we study the average and typical spin correlations to characterize this disordered phase.
We analyze the average spin correlation in two ways.
First, we consider the $L_y$-dependence, which has been used in the study of the $J-Q$ model~\cite{liu2018}.
We calculate the average absolute spin correlations that are defined for the largest-distance sites on the middle $L_y \times L_y$ subsystem, as illustrated in the inset of Fig.~\ref{fig:cor}(a).
The log-log plot of the average spin correlation versus $L_y$ shows a power-law decay $C_{s}(L_y) \propto L_{y}^{-\alpha}$ with the exponent $\alpha \simeq 2$.
This exponent agrees with that found in the RS state of the $J-Q$ model~\cite{liu2018}.
Secondly we study the average spin correlation decay along the $x$ direction.
To reduce finite-size effects, we consider the large systems with $L_y = 8, 10$ and $L_x = 2 L_y$.
We choose the reference site ${\bf S}_0$ on the left side of the middle $L_y \times L_y$ subsystem and study the spin correlation decay $\left [ \left | \left \langle {\bf S}_{0}\cdot {\bf S}_{r} \right \rangle \right | \right ]$ from the left to the right side, as shown in Fig.~\ref{fig:cor}(b).
The average spin correlations shown here are very close, showing small finite-size effects and the consistent power-law decay $C_{s}(r) \propto r^{-2}$ (see SM for the smaller-size data and the consistent results for the weaker randomness strength, as well as the preliminary results of the dimer-dimer correlations~\cite{suppl}).

We further investigate the probability distribution of the absolute spin correlations defined for the largest-distance sites in the middle $L_y \times L_y$ subsystem, denoted as $|C_{ij}|$.
The probability $P(|C_{ij}|)$ versus $\ln |C_{ij}|$ on different sizes becomes broader with growing $L_y$~\cite{suppl}, which agrees with the decrease of $C_{s}(L_y)$ .
To eliminate the finite-size effects in the analysis, we define a new scaling variable $x = \ln ( |C_{ij}| L^{2}_{y})$ and plot $P(x)$ versus $x$ in Fig.~\ref{fig:cor}(c-d).
Any distribution $P(x)$ with data collapse for different $L_y$ must lead to the power-law decay of the average spin correlation
\begin{equation}
C_{s}(L_y) = \int_{0}^{\infty} d x P(x) \frac{e^x}{L^{2}_y} \propto L^{-2}_{y}.
\end{equation}
Thus, the good data collapse in Fig.~\ref{fig:cor}(c-d) further supports the power-law decay of $C_{s}(L_y)$ found in Fig.~\ref{fig:cor}(a).

The logarithmically broad distribution of spin correlations may lead to the exponential decay of the typical spin correlation defined as $\exp{[\ln |\langle {\bf S}_0 \cdot {\bf S}_r \rangle|]}$. In the 1d RS state, the exponential typical spin correlation can be equivalently described as $- [\ln |\langle {\bf S}_0 \cdot {\bf S}_r \rangle|] \propto r^{1/2}$~\cite{fisher1994}.
We show the linear plot of $- [\ln |\langle {\bf S}_0 \cdot {\bf S}_r \rangle|]$ versus $r^{1/2}$ in Fig.~\ref{fig:cor}(e-f).
On both the $L_y = 8, 10, L_x = 2 L_y$ systems and the $L_y = 4, 6$ cylinders with large $L_x$, $- [\ln |\langle {\bf S}_0 \cdot {\bf S}_r \rangle|]$ follows the linear scaling with $r^{1/2}$ quite well.
We also analyze $- [\ln |\langle {\bf S}_0 \cdot {\bf S}_r \rangle|]$ versus $r$ using the log-log manner~\cite{suppl}, which shows good power-law dependence and the slope gives the power exponent very close to $1/2$, supporting the results in Fig.~\ref{fig:cor}(e-f).
Therefore, our results unveil the characteristic behaviors of spin correlations, which suggest the RS nature of this disordered phase. 

\begin{figure}[htb]
\includegraphics[width=0.239\textwidth]{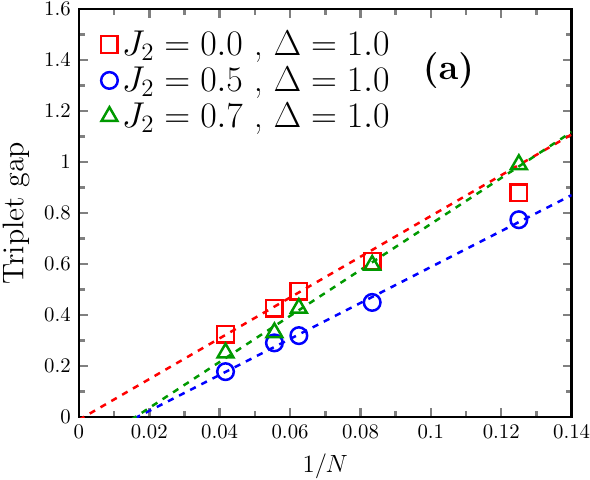}
\includegraphics[width=0.239\textwidth]{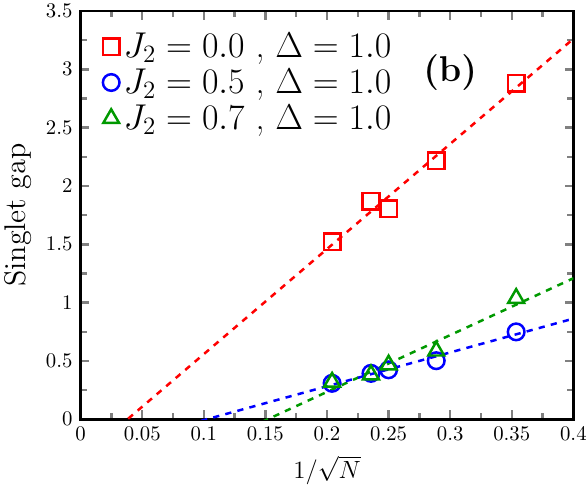}
\includegraphics[width=0.48\textwidth]{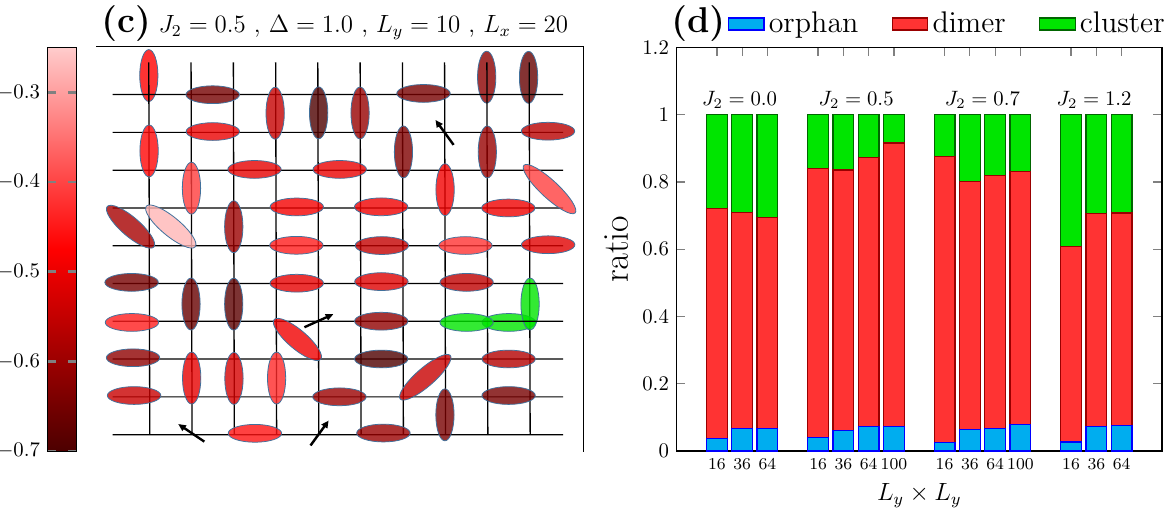}
\caption{Finite-size scaling of gaps and distributions of microscopic clusters. (a) and (b) are the gaps obtained from the ED calculation. $N = 8, 12, 16, 18, 24$ is the total site number. The linear fitting curves are guides to the eye. (c) The covering of the isolated dimers (the red ellipse), the resonating-dimer clusters (the connected green ellipses), and the orphan spins (the arrow) on the lattice with a given random sample. The gradation of the red color corresponds to the strength of the bond correlation $\langle {\bf S}_i \cdot {\bf S}_j \rangle$.
(d) The ratios of the different clusters in different quantum phases. We analyze the data from the middle $L_y \times L_y$ subsystem.}
\label{fig:gap}
\end{figure}

{\it Excitation gaps and microscopic clusters.}
The previous ED study found the vanished spin gap in the disordered phase~\cite{uematsu2018}.
To further unveil the properties of excitations, we separately compute spin-triplet and singlet gaps by using ED.
Compared with the results in the N\'eel phase, both triplet and singlet gaps in the disordered phase decrease, as shown in Fig.~\ref{fig:gap}(a-b), showing the vanished gaps in the thermodynamic limit~\cite{note1}.
In particular, while the finite-size data of triplet gap slightly decreases, the singlet gap is strongly suppressed.

In the presence of disorder, local clusters may form in microscopic scale.  
Recent ED study has found three types of clusters in random spin systems, including the isolated dimer, the resonating-dimer cluster, and the orphan spin~\cite{kawamura2019}, which should be helpful for understanding low-energy excitations.
We extend such analysis in different phases based on the DMRG results.
For each sample on the $L_x = 2 L_y$ cylinder, we collect all the correlations $\langle {\bf S}_i \cdot {\bf S}_j \rangle$ in the middle $L_y \times L_y$ subsystem.
We define the isolated dimer as the singlet pair with strong correlation, the resonating-dimer cluster as the cluster with more than two strongly-coupled spins, and the orphan spin as the one weakly coupled to other spins (see SM~\cite{suppl} and Ref.~\onlinecite{kawamura2019} for the procedure to obtain the covering of clusters).
One example of the covering is shown in Fig.~\ref{fig:gap}(c).
For each random sample we compute the ratios for the different types of clusters and then take their sample averages, as shown in Fig.~\ref{fig:gap}(d).
In the three quantum phases, the dominant cluster is always the isolated dimer due to strong bond randomness, and the ratio of orphan spin varies slightly.
A feature in the disordered state is the less resonating dimers compared with magnetic states.
In the RS picture, the triplet excitations are related to breaking dimers.
The broad distribution of bond couplings cause a decrease of the triplet excitation gap.
For singlet excitations which are related with spin flip in magnetic phases, in the RS state such excitations could be driven by the diffusion of orphan spins~\cite{kawamura2019}, which may account for the strongly suppressed singlet gap in Fig.~\ref{fig:gap}(b).
Due to the size limit, here we only focus on the features of the small clusters. The collective properties on the larger length scale need future study beyond the present system size.

{\it Conclusion and discussion.} We have studied the ground states of the spin-$1/2$ square Heisenberg model with random $J_1, J_2$ couplings in the strong disorder regime by using the large-scale DMRG calculation.
With growing $J_2$, we find a gapless phase without any magnetic order or spin glass order.
We identify the characteristic size-scaling behaviors of spin correlations in this state, including the $L^{-1/2}_x$ scaling of spin-freezing parameter with system length $L_x$, the $r^{-2}$ power-law decay of the average spin correlation, and the exponential decay of the typical spin correlation.
Although the RS state of a generic 2d system is not well understood, our results strongly suggest this disordered state as the 2d analog of the 1d RS state.
The same scaling behavior of average spin correlation in this frustrated model and the $J-Q$ model also suggests the same fixed point of the states in these different systems. 
For further study, the dynamic exponent may be explored by the tensor network simulation at finite temperature~\cite{chen2018, li2020} or the analysis of excitation gap distribution (a rough estimation of dynamic exponent from ED gap data is shown in SM~\cite{suppl}).

For Sr$_2$CuTe$_{1-x}$W$_{x}$O$_6$, spin freezing has been excluded for $x = 0.5$~\cite{mustonen2018}, which agrees with our results.
For more mixing ratios, the new $\mu$SR measurement has identified a large dynamic exponent supporting the intermediate RS phase in Sr$_2$CuTe$_{1-x}$W$_{x}$O$_6$~\cite{hong2020}.
At last, we emphasize that this 2d RS state should be considered as the preliminary understanding of the spin-liquid-like state in Sr$_2$CuTe$_{1-x}$W$_{x}$O$_6$, since the bond randomness in our study is a simplified picture to describe the random substitution in Sr$_2$CuTe$_{1-x}$W$_{x}$O$_6$~\cite{hong2020}.
We leave the more realistic modeling of the material to future study.

{\it Acknowledgement.} We acknowledge the discussions with Hikaru Kawamura, Victor Quito, Yu-Cheng Lin, and Anders Sandvik. This work was supported by the National Natural Science Foundation of China grants 11874078, 11834014, grant 11804401, and the Fundamental Research Funds for the Central Universities. The work at CSUN was supported by the U.S. Department of Energy, Office of Basic Energy Sciences under the grant No. DE-FG02-06ER46305 (D.N.S.).

\bibliography{random}

\begin{thebibliography}{76}
\bibitem [{not()}]{note1}%
  \BibitemOpen
  \href@noop {} {}\bibinfo {note} {Since the N\'eel and stripe
  antiferromagnetic phase break SU(2) symmetry spontaneously in the
  thermodynamic limit, the energy spectrum has the Anderson tower of states on
  finite-size systems. With growing system size, the tower of states levels
  scale with $1/N$ to the ground state, and the low-energy magnon excitations
  scale with $1/\sqrt{N}$, where $N$ is the total number of sites. Therefore,
  we scale the triplet gap with $1/N$ and singlet gap with
  $1/\sqrt{N}$.}\BibitemShut {Stop}%
\end{thebibliography}

\begin{center}
\begin{Large}
{Supplemental Material}
\end{Large}
\end{center}

\section{Randomness sample dependence of the results and the DMRG convergence}

\begin{figure}[htb]
\includegraphics[width=0.4\textwidth]{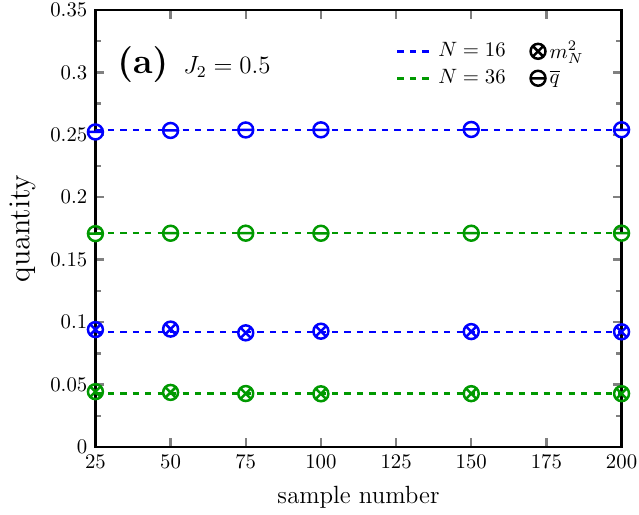}
\includegraphics[width=0.4\textwidth]{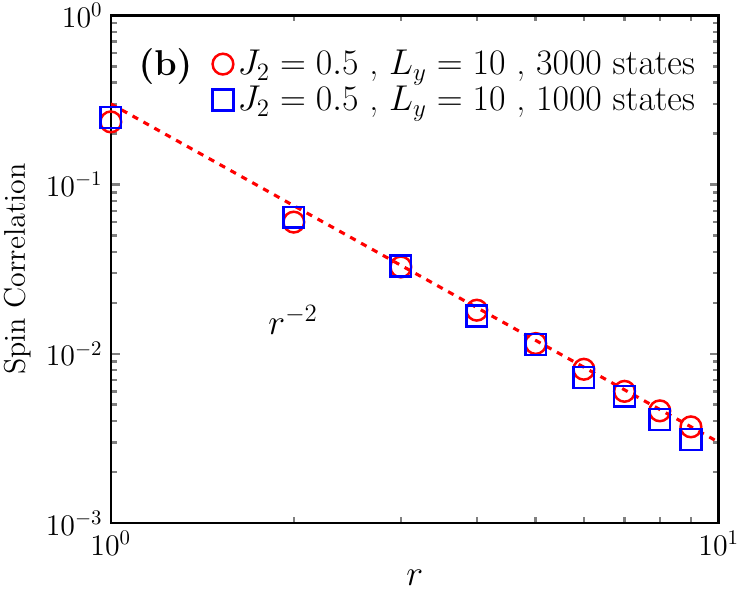}
\caption{Convergence of the DMRG simulation results.
(a) Randomness sample dependence of the physical quantities.
We take $J_2 = 0.5, \Delta = 1.0, L_y = 4, 6, L_x = 2 L_y$. We show the spin-freezing parameter $\overline{q}$ and N\'eel order parameter $m^{2}_{\rm N}$ obtained from the middle $L_y \times L_y$ subsystem and the sample number $25 - 200$. (b) Bond dimension dependence of the average spin correlation function along the axis direction of the middle $L_y \times L_y$ subsystem on the $L_y = 10$ cylinder. The data are obtained for $J_2 = 0.5, \Delta = 1.0$ by keeping $1000$ and $3000$ SU(2) states.}
\label{fig:convergence}
\end{figure}

To compute the physical quantities of the system with randomness, one needs to take the sample average for different randomness samples.
In the exact diagonalization (ED) calculation, we use the randomness samples up to $200 - 500$.
In the density matrix renormalization group (DMRG) simulation, we could not take so many samples because of the expensive computation time.
To estimate the randomness sample dependence of the physical quantities, we have tested the average results calculated from different numbers of randomness samples.
The sample dependences of the spin-freezing parameter $\overline{q}$ and the N\'eel order parameter $m^{2}_{\rm N}$ for $J_2 = 0.5, \Delta = 1.0$ are shown in Fig.~\ref{fig:convergence}(a) on the $L_y = 4, 6$ cylinders.
We take the number of randomness samples from $25$ to $200$, and one can see that both quantities converge very fast with the growing sample number. 
Therefore, in our DMRG simulation we take $100$ and $50$ randomness samples for $L_y = 4, 6$ and $L_y = 8, 10$ cylinders respectively, which ensure the good sample average of the quantities.

In Fig.~\ref{fig:convergence}(b), we also compare the average spin correlations for $J_2 = 0.5, \Delta = 1.0$ on the $L_y = 10$ cylinder. With the growing bond dimensions from $1000$ to $3000$ SU(2) states, the average spin correlations only increase slightly, which indicates that our demonstrated results by keeping $3000$ states have relatively good convergence.

\section{Analysis of the stripe magnetic order}

\begin{figure}[htb]
\includegraphics[width=0.4\textwidth]{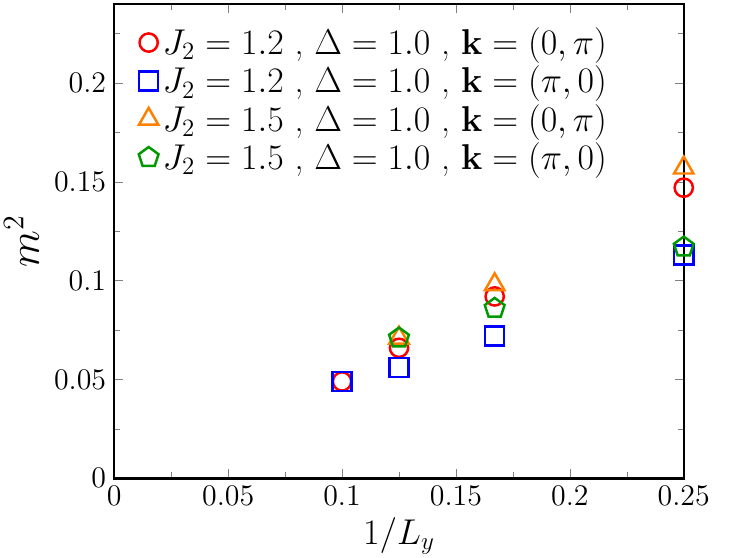}
\caption{Magnetic order parameter for $J_2 = 1.2, 1.5$ and $\Delta = 1.0$. We show the results at ${\bf k} = (0,\pi)$ and $(\pi,0)$.}
\label{supplfig:stripe}
\end{figure}

In Fig.~2(b) of the main text, we have shown the finite-size scaling of the stripe magnetic order $m^{2}_{str}$, where we have taken the average value of $m^{2}(0,\pi)$ and $m^{2}(\pi,0)$ as $m^{2}_{str}$.
Here we show $m^{2}(0,\pi)$ and $m^{2}(\pi,0)$ separately for $J_2 = 1.2, 1.5$ and $\Delta = 1.0$ in Fig.~\ref{supplfig:stripe}.
While the data of $m^{2}(0,\pi)$ may be extrapolated to zero in the thermodynamic limit, the results of $m^{2}(\pi,0)$ clearly will go to finite, which can also characterize the emergent stripe magnetic order.

\section{Spin correlation square and scaling behavior of spin-freezing parameter}

\begin{figure}[htb]
\includegraphics[width=0.4\textwidth]{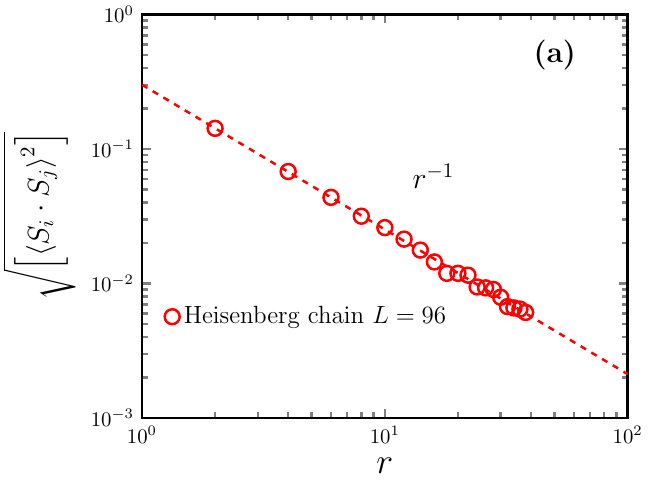}
\includegraphics[width=0.4\textwidth]{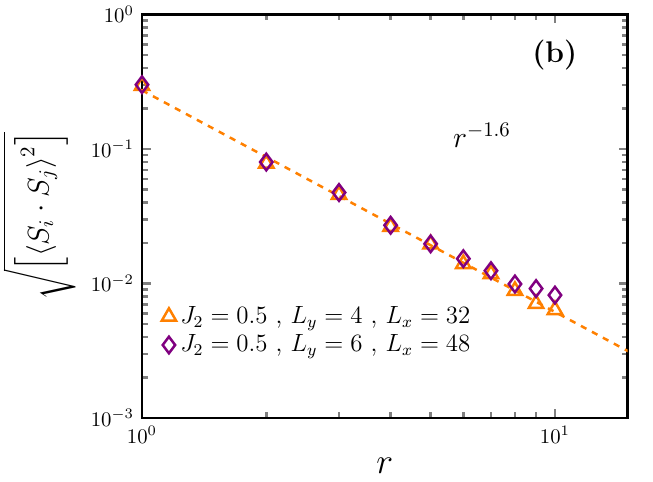}
\includegraphics[width=0.4\textwidth]{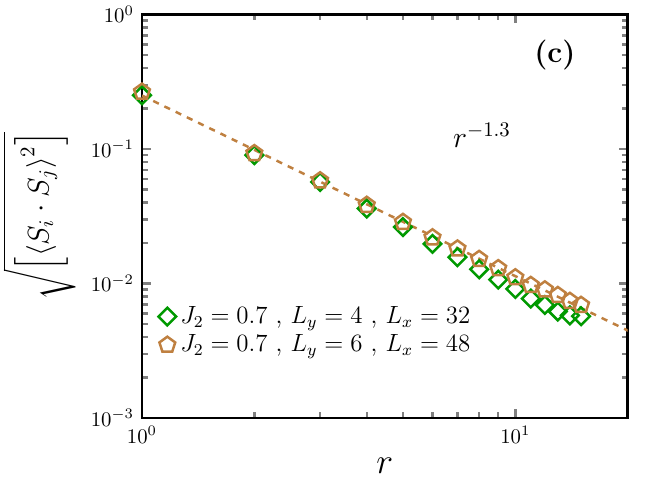}
\caption{Log-log plot of the average spin correlation square versus the distance. $r$ denotes the distance of the two sites $|i-j|$. (a) is the result of the RS state in the 1d Heisenberg chain with total site number $L = 96$. (b) and (c) are the results in the RS state with $J_2 = 0.5, 0.7$ and $\Delta = 1.0$ on the $L_y = 4, 6$ cylinders. The average spin correlation square shows a power-law decay in these systems with different power exponents.}
\label{fig:spin_square}
\end{figure}

In the main text, we have shown the $L^{-1/2}_{x}$ scaling of the spin-freezing parameter $\overline{q}$ for the given $L_y$ systems in the one-dimensional (1d) random-singlet (RS) state of the random Heisenberg chain ($L_y = 1$) and the RS state in the two-dimensional (2d) frustrated square model on finite cylinder.
The spin-freezing parameter $\overline{q}$ is defined as
\begin{equation} \label{suppleq:freezing}
\overline{q} = \frac{1}{N_s} \sqrt{\sum_{i, j} \left[ \langle {\bf S}_{i} \cdot {\bf S}_{j} \rangle^2 \right]},
\end{equation}
where $N_s = L_x \times L_y$ and the summation runs over all the $N_s$ sites.
``$\langle \rangle$" and ``[ ]" are used to represent quantum mechanical and stochastic averages, respectively.
For the given $L_y$ system, the size scaling of $\overline{q}$ with $L_x$ is determined by the decay behavior of $\left[ \langle {\bf S}_{i} \cdot {\bf S}_{j} \rangle^2 \right]$ versus $|i - j|$ (we denote $|i - j| \equiv r$).
When computing the scaling behavior of $\overline{q}$ along one direction (like a 1d scaling), if we take the spin correlation decay as $\sqrt{\left[ \langle {\bf S}_{i} \cdot {\bf S}_{j} \rangle^2 \right]} \sim r^{-a}$, then we can find that $\overline{q} \sim L^{-1/2}_x$ if the power exponent $a > 1/2$.

In Fig.~\ref{fig:spin_square}, we show the log-log plots of $\sqrt{\left[ \langle {\bf S}_{i} \cdot {\bf S}_{j} \rangle^2 \right]}$ versus $r$ in the RS phases.
In the 1d RS phase of the Heisenberg chain, our DMRG result shows that $\sqrt{\left[ \langle {\bf S}_{i} \cdot {\bf S}_{j} \rangle^2 \right]} \sim r^{-1}$, which naturally leads to the $L^{-1/2}_x$ scaling behavior of $\overline{q}$.
In the RS phase of the frustrated square-lattice model, our data show that $\sqrt{\left[ \langle {\bf S}_{i} \cdot {\bf S}_{j} \rangle^2 \right]}$ also follows the algebraic decay with the power exponent $\sim -1.6$ and $-1.3$ for $J_2 = 0.5$ and $J_2 = 0.7$, respectively.
These exponents also satisfy the relation $a > 1/2$ and thus support the $L^{-1/2}_x$ scaling behavior of $\overline{q}$, as we have shown in the main text.

\section{Power-law decay of average spin correlation}

In the main text, we have shown the power-law decay of the average spin correlation in the RS phase on the $L_y = 8, 10, L_x = 2 L_y$ cylinders.
Here we demonstrate the results on the $L_y = 4, 6$ cylinders with the much larger $L_x$ in Fig.~\ref{fig:cor}.
Following the definition in the main text, we study the average spin correlation decay along the $x$ direction in the bulk of cylinder.
For the $L_y = 4, L_x = 32$ system, the average spin correlations also agree with the decay behavior $C_s(r) \propto r^{-2}$.  
On the $L_y = 6, L_x = 48$ cylinder, the average spin correlations seem to follow the power exponent $-1.8$, which is also close to the $r^{-2}$ behavior.
For $J_2 = 0.7$, the exponent $\sim -2$ at long distance although it slightly deviates at short distance, similar to our results for $L_y = 8, 10$ shown in the main text.

\begin{figure}[htb]
\includegraphics[width=0.4\textwidth]{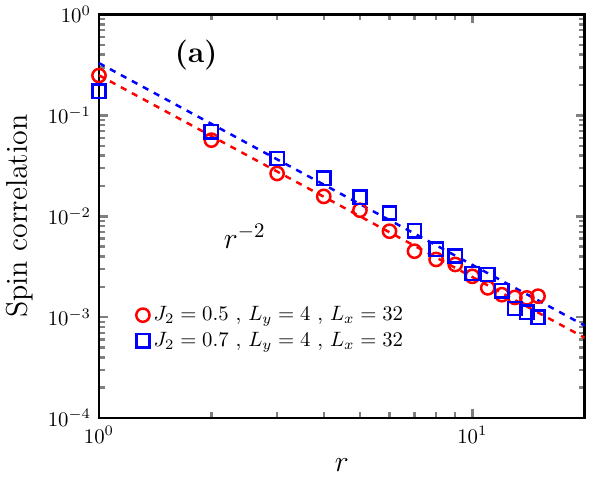}
\includegraphics[width=0.4\textwidth]{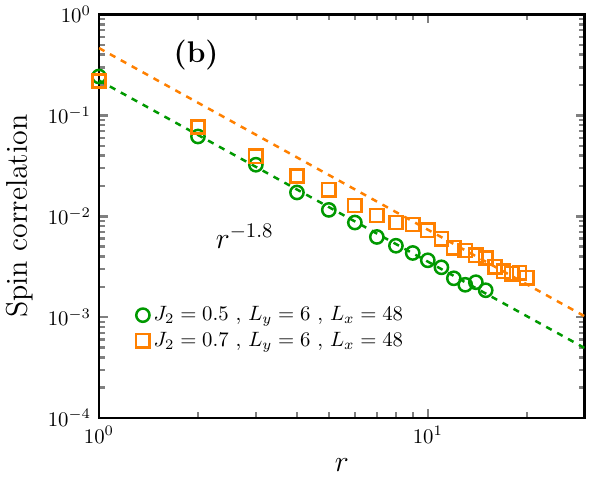}
\includegraphics[width=0.4\textwidth]{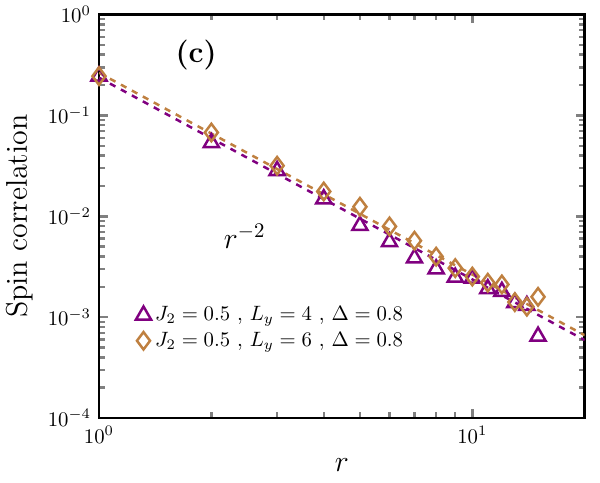}
\caption{Power-law decay of average spin correlation function along the axis direction of cylinder.
$J_2 = 0.5, 0.7, \Delta = 1.0$ for (a) $L_y = 4$ and (b) $L_y = 6$. The $L_y = 4$ and $L_y = 6$ data are obtained on the $L_x = 32$ and $L_x = 48$ cylinders, respectively. (c) $J_2 = 0.5, \Delta = 0.8$ on the $L_y = 4, 6$ cylinders.
}
\label{fig:cor}
\end{figure}

\begin{figure}[htb]
\includegraphics[width=0.4\textwidth]{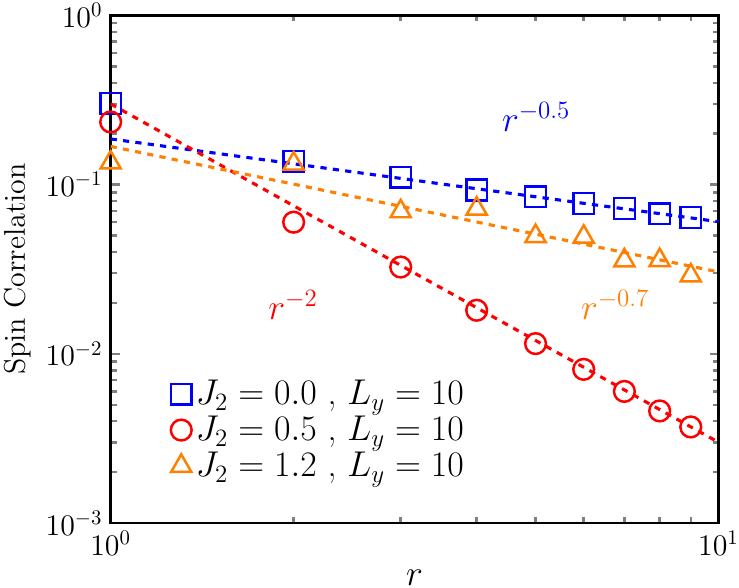}
\caption{Comparing the average spin correlations in the different quantum phases. We choose three parameters on the $L_y = 10$ cylinder as representatives: $J_2 = 0.0$ for the N\'eel phase, $J_2 = 1.2$ for the stripe order phase, and $J_2 = 0.5$ for the disordered phase. The data are shown in the double-logarithmic scale.}
\label{fig:cor_copar}
\end{figure}

In the main text, we mainly focus on the strong randomness case with $\Delta = 1.0$.
For checking the universality of this $r^{-2}$ power-law decay of average spin correlation in the RS phase, we have also considered the systems with the smaller randomness strength $\Delta$, which indeed also show the consistent $r^{-2}$ decay behavior.
One example for $J_2 = 0.5, \Delta = 0.8$ on the $L_y = 4, 6$ systems is shown in Fig.~\ref{fig:cor}(c), which supports the universality of this $r^{-2}$ behavior of the average spin correlation decay in the RS phase.

In the N\'eel and stripe magnetic order phase, spin correlation functions decay much more slowly compared with those in the RS phase. In Fig.~\ref{fig:cor_copar}, we show the average spin correlations in the three different quantum phases on the $L_y = 10$ cylinder. All the data are obtained on the middle $L_y \times L_y$ subsystem and are shown in the double-logarithmic scale. 
We roughly fit the data algebraically, which finds the much smaller power exponents in the magnetic order phases.
The strong average spin correlations at $J_2 = 1.2$ also agree with the emergent stripe magnetic order.

\section{Analysis of the dimer-dimer correlation}

In the 1d RS phase, the average dimer-dimer correlation function follows the $1/r^4$ algebraic decay.
In the study of the random 2d $J-Q$ model~\cite{liu2018}, the average dimer correlation functions have also been calculated.
Although the long-distance dimer correlations have relatively large error bar, the intermediate-distance data seem to follow the $1/r^4$ algebraic decay.

\begin{figure}[htb]
\includegraphics[width=0.4\textwidth]{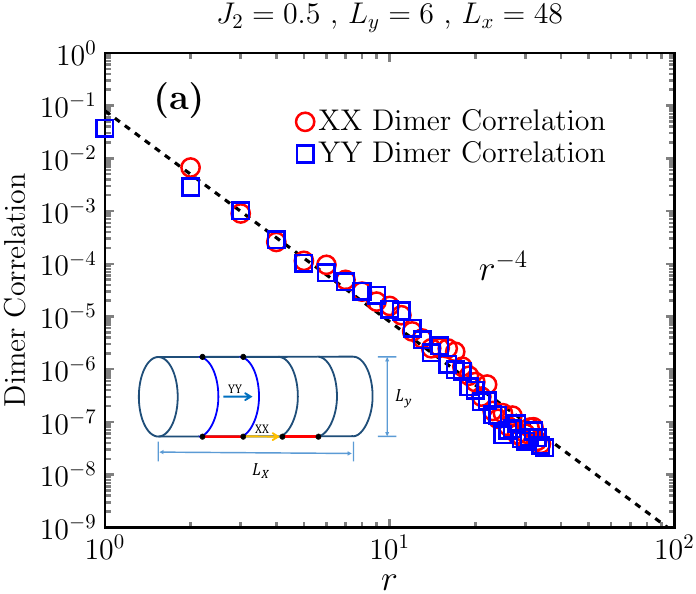}
\includegraphics[width=0.4\textwidth]{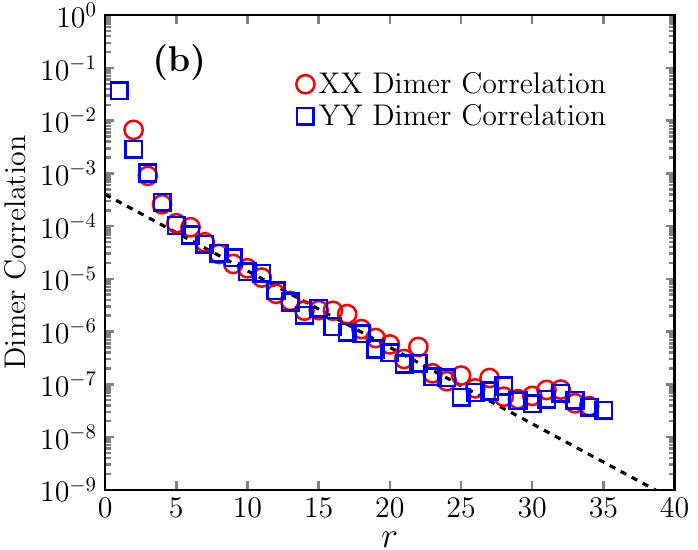}
\caption{Average dimer correlation functions in the intermediate RS phase. We demonstrate the results for $J_2 = 0.5, \Delta = 1.0$ on the $L_y = 6$ $L_x = 48$ cylinder as an example. The inset in (a) shows the definitions of the XX dimer and the YY dimer correlations, which measure the dimer correlations between the horizontal bonds and the vertical bonds, respectively. The reference bonds are chosen in the seventh column to avoid edge effect. The data are demonstrated as the double-logarithmic scale in (a) and the semi-logarithmic scale in (b). The dashed line in (a) denotes the $1/r^{4}$ algebraic fitting for most of the data.}
\label{fig:dimer_cor}
\end{figure}

In our DMRG calculation, we also investigate the dimer correlation function in the RS phase. For two nearest-neighbor bonds $i,j$ and $k,l$, the average dimer correlation is defined as
\begin{equation}
\left[ \langle ({\bf S}_i \cdot {\bf S}_j)({\bf S}_k \cdot {\bf S}_l) \rangle - \langle {\bf S}_i \cdot {\bf S}_j \rangle \langle {\bf S}_k \cdot {\bf S}_l \rangle \right].
\end{equation}
We define the XX (YY) dimer correlations to measure the correlations between the horizontal (vertical) bonds, as illustrated by the inset of Fig.~\ref{fig:dimer_cor}(a).
In Fig.~\ref{fig:dimer_cor}, we show the average dimer correlations for $J_2 = 0.5, \Delta = 1.0$ on the $L_y = 6, L_x = 48$ cylinder.
We choose the reference bonds in the seventh column to avoid edge effect.
For computing both the XX and YY dimer correlations, we also average the results along the six rows. 
In the double-logarithmic plot Fig.~\ref{fig:dimer_cor}(a), the two types of average dimer correlations are close and both of them appear to follow the algebraic decay. 
We use the $r^{-4}$ decay to fit the results, which can describe most of the data, indicating that our results are consistent with the $r^{-4}$ decay.
We also demonstrate the same data as the semi-logarithmic scale in Fig.~\ref{fig:dimer_cor}(b).
It seems that for the distance $r =5 - 25$, the results can also be fitted by the exponential decay. 
Although it is hard to pin down the decay behavior for such rapidly decaying correlations, our results could be still consistent with the $r^{-4}$ decay, which further supports the similar RS nature of this disordered state to the 1d RS state.
Since the four-site dimer correlations are more challenging to converge, we only demonstrate the results for $L_y = 6$ here and we leave the calculations on the larger system size to further study.

\section{Probability distribution of spin correlation}

\begin{figure}[htb]
\includegraphics[width=0.4\textwidth]{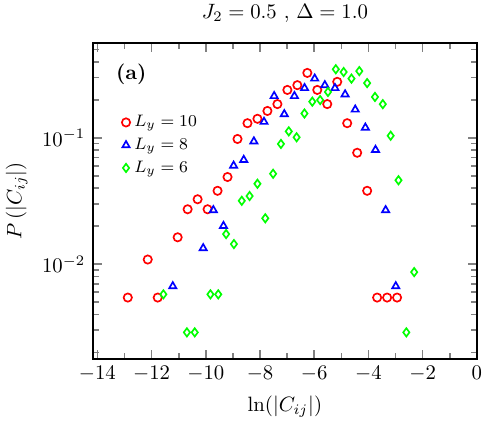}
\includegraphics[width=0.4\textwidth]{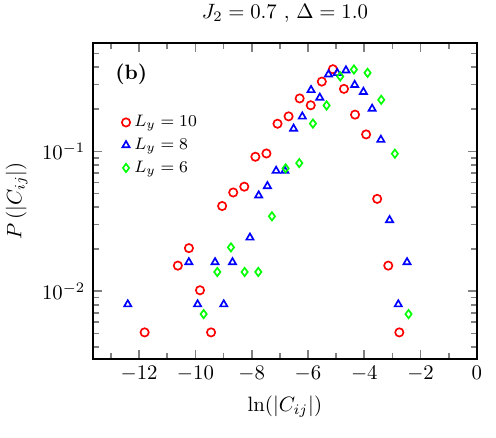}
\caption{Histograms showing the probability distribution of spin correlation. $P(|C_{ij}|)$ versus $\ln(|C_{ij}|)$ with $C_{ij}$ denoting the spin correlation of the largest-distance sites on the middle $L_y \times L_y$ subsystem. The data include the system sizes $L_y = 6, 8, 10$ for (a) $J_2 = 0.5, \Delta = 1.0$ and (b) $J_2 = 0.7, \Delta = 1.0$.}
\label{fig:pp}
\end{figure}

In the main text, we have analyzed the $L_y$ dependence of the average spin correlation $C_{s}(L_y)$ that is defined for the largest-distance sites on the middle $L_y \times L_y$ subsystem of cylinder.
The size scaling shows the $C_{s}(L_y) \propto L^{-2}_{y}$ behavior in the RS phase, which is further supported by the data collapse of the probability distribution $P(x)$ with the variable $x = \ln(|C_{ij}| L^2_y)$, where $C_{ij}$ is the spin correlation $\langle {\bf S}_i \cdot {\bf S}_j \rangle$ of the largest-distance sites.
Here we directly show the probability distribution $P(|C_{ij}|)$ versus $\ln |C_{ij}|$ in Fig.~\ref{fig:pp}.
We can see that $P(|C_{ij}|)$ becomes broader with growing $L_y$, which is consistent with the decreased spin correlation $C_{s}(L_y)$ with growing $L_y$ in the studied RS phase.

\begin{figure}[htb]
\includegraphics[width=0.4\textwidth]{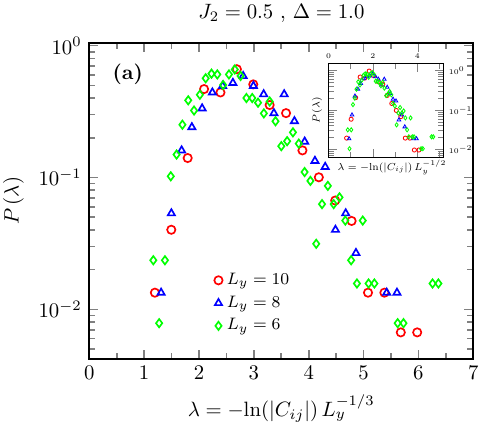}
\includegraphics[width=0.4\textwidth]{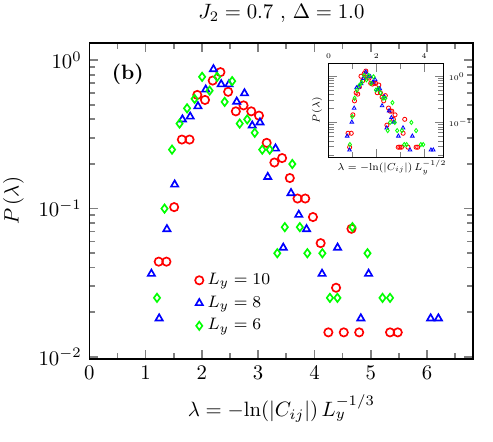}
\caption{Histograms showing the probability distribution of spin correlation. $P(\lambda)$ versus $\lambda = - \ln (|C_{ij}|) L^{-\beta}_y$ with $\beta = 1/3$ and $1/2$. $C_{ij}$ defines the spin correlation of the largest-distance sites on the middle $L_y \times L_y$ subsystem. (a) and (b) show the data collapse for $J_2 = 0.5, 0.7, \Delta = 1.0$ on the system sizes $L_y = 6, 8, 10$. }
\label{fig:prodistri}
\end{figure}

Before using the new variable $x = \ln(|C_{ij}| L^2_y)$, we have also considered the regular scaling method to analyze the probability distribution, which however has a large uncertainty due to the limit of our system size.
For this reason, we define the new variable $x = \ln(|C_{ij}| L^2_y)$, which has no adjustable parameter and any data collapse would support the $C_{s}(L_y) \propto L^{-2}_{y}$ behavior.
Here, we show the uncertainty of the regular analysis.
Following the general idea, we define a new scaling variable $\lambda$ to study the size scaling of the probability distribution, 
\begin{equation}
\lambda = - ( \ln |C_{ij}| ) L_y^{-\beta}, \label{eq:lambda}
\end{equation}
and transform the histograms from the distribution $P(|C_{ij}|)$ to $P(\lambda)$.
With the correct exponent $\beta$, the probability distributions $P(\lambda)$ for different system sizes would collapse together.
Meanwhile, the average spin correlation $C_{s}(L_y)$ obtained from the integral of the data-collapsed distributions
\begin{equation}
C_{s}(L_y) = \int_{0}^{\infty} d \lambda e^{- \lambda L^{\beta}_y } P(\lambda)
\end{equation}
should agree with the direct calculation of $C_{s}(L_y)$.
We have tried different $\beta$ to collapse the data for $L_y = 6 - 10$.
The results with $\beta = 1/3$ and $1/2$ are shown in Fig.~\ref{fig:prodistri}.
One may find that for both $\beta$ values, the correlation distributions collapse equally well, especially in the small $\lambda$ regime where spin correlations are large.
By tuning $\beta$, we find that the data collapse seems good in our resolution for the range of $0.3 \lesssim \beta \lesssim 0.8$. 
In the RS phase of the random $J - Q$ model on the square lattice, $\beta$ is found to be $1/3$ by using the data collapse of probability distribution with system size up to $48 \times 48$~\cite{liu2018}, which is in the range of the estimated $\beta$ in this $J_1 - J_2$ model.
We expect that the future study on the larger system size can obtain more accurate scaling exponent $\beta$ for the RS state in frustrated spin systems.

\begin{figure*}[htb]
\includegraphics[width=0.4\textwidth]{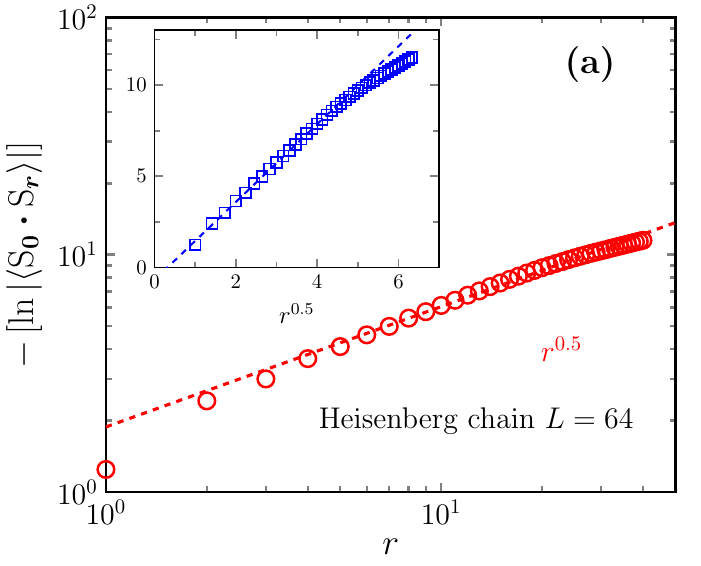}
\includegraphics[width=0.4\textwidth]{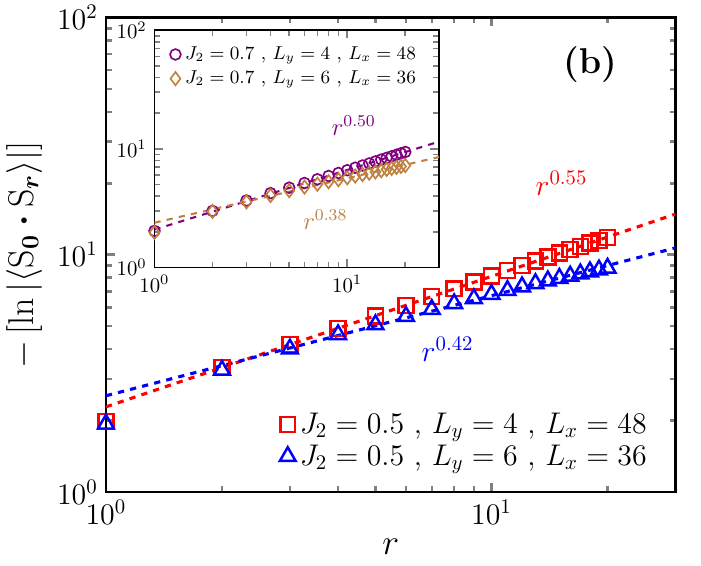}
\includegraphics[width=0.4\textwidth]{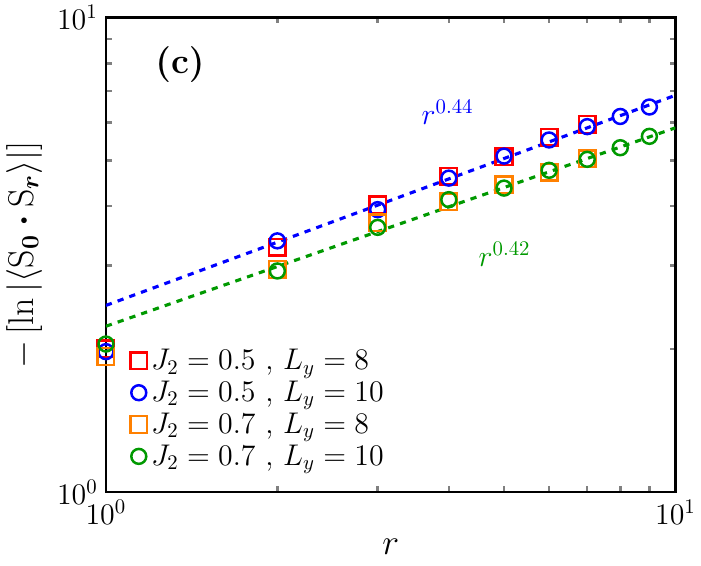}
\includegraphics[width=0.4\textwidth]{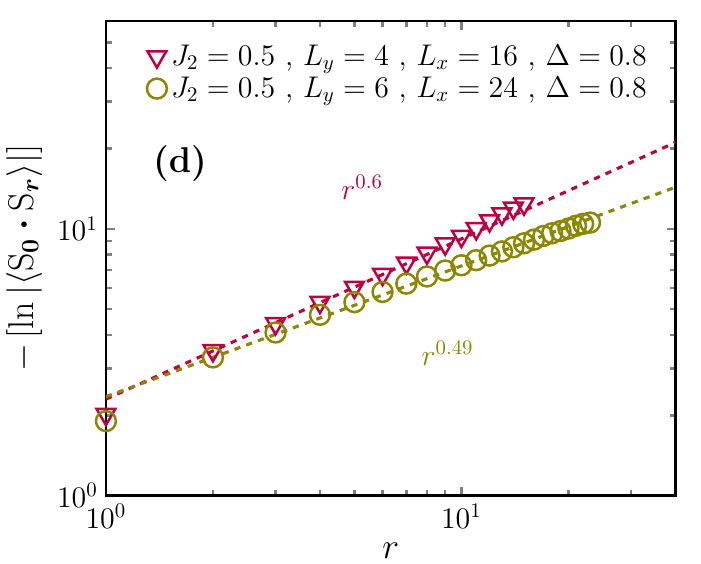}
\caption{Power-law dependence of $- [\ln |\langle {\bf S}_0 \cdot {\bf S}_r \rangle|]$ as a function of $r$ in the RS phase. (a) Log-log plot of the result for the 1d Heisenberg chain with random couplings. The inset shows the same data as a function of $r^{1/2}$ in the linear plot. (b) and (c) are the log-log plots of the results on the $L_y = 4, 6$ systems with long $L_x$ and the $L_y = 8, 10, L_x = 2 L_y$ systems, respectively. (d) Log-log plot of the results on the systems with the randomness strength $\Delta = 0.8$ on the $L_y = 4, 6$ systems.
}
\label{fig:typcor}
\end{figure*}

\section{Exponential decay of typical spin correlation}

In the main text, we have shown $- [\ln |\langle {\bf S}_0 \cdot {\bf S}_r \rangle|]$ as a function of $r^{1/2}$ in the RS state, which is proportional to $r^{1/2}$ and thus supports the exponential decay of typical spin correlation~\cite{fisher1994}.
Here, we take another way to analyze the data by plotting $- [\ln |\langle {\bf S}_0 \cdot {\bf S}_r \rangle|]$ as a function of the distance $r$ in the log-log manner.

First of all, we show our result of the 1d Heisenberg chain with random couplings in Fig.~\ref{fig:typcor}(a).
We plot $- [\ln |\langle {\bf S}_0 \cdot {\bf S}_r \rangle|]$ as a function of either $r$ (log-log plot) or $r^{1/2}$ (linear plot), which consistently support the predicted behavior $- [\ln |\langle {\bf S}_0 \cdot {\bf S}_r \rangle|] \propto r^{1/2}$ by the strong-disorder renormalization group~\cite{fisher1994}.
In Fig.~\ref{fig:typcor}(b-c), we demonstrate $- [\ln |\langle {\bf S}_0 \cdot {\bf S}_r \rangle|]$ as a function of $r$ in the RS phase of the studied square-lattice model.
We can find that $- [\ln |\langle {\bf S}_0 \cdot {\bf S}_r \rangle|]$ fits the power-law dependence with the distance $r$, which also indicates the exponential decay of the typical spin correlation in this 2d RS state.
The fitted power exponents change slightly with growing $L_y$ and seem to near convergence on the $L_y = 8, 10$ systems, which are also near $1/2$ and are consistent with our analyses in the main text.

Furthermore, we also demonstrate the similar analysis of the results at the randomness strength $\Delta = 0.8$ in Fig.~\ref{fig:typcor}(d).
Here we have only studied the systems with $L_y = 4, 6$.
We find that $- [\ln |\langle {\bf S}_0 \cdot {\bf S}_r \rangle|]$ also follows a power-law dependence with $r$. 
For $L_y = 4$, the power exponent is about $0.6$.
However, on the wider $L_y = 6$ system, the exponent rapidly decreases to $0.49$, very close to $1/2$.
Our results indicate that not only the average spin correlation but also the typical spin correlation have the universal behavior in the RS phase.

\section{Algebraic extrapolation of excitation gaps}

In Fig. 4(a-b) of the main text, we have shown the system size dependence of the spin gaps obtained by ED, which demonstrate the vanished triplet and singlet gaps in the RS phase. 
Here, we roughly estimate the dynamic exponent $z$ from the ED data of the gaps.
In a critical phase, the asymptotic behavior of excitation gap scales as $1/L^z$, where we choose $N = L \times L$ as the total number of cluster sites.
In our study, since the ED results strongly suffer from finite-size effect, we choose to fit the ED data using algebraic extrapolation and estimate the dynamic exponent from the fitting.

\begin{figure}[htb]
\includegraphics[width=0.4\textwidth]{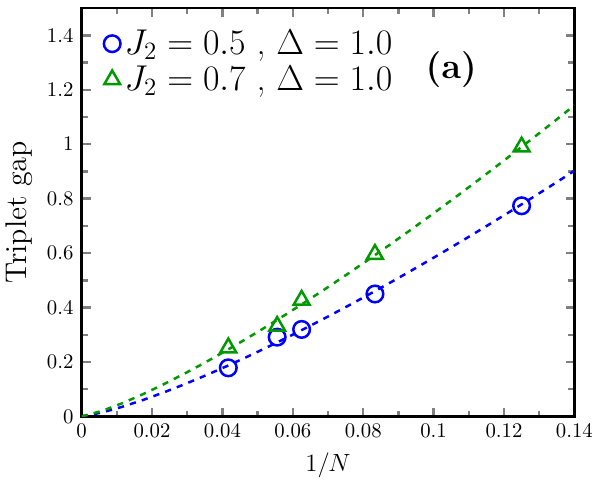}
\includegraphics[width=0.4\textwidth]{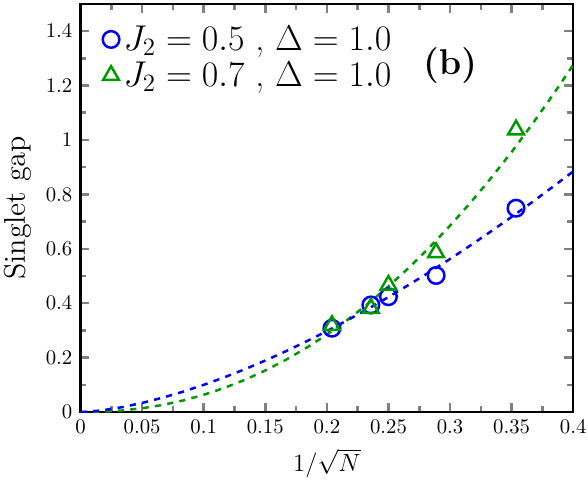}
\caption{Algebraic extrapolation of triplet and singlet gaps in the RS phase. The gap data are the same as those shown in Fig. 4(a-b) of the main text, obtained from ED calculation. The dashed lines in (a) show the fittings of triplet gap $\sim N^{-\alpha}$, which give the dynamic exponent $z = 2\alpha$. We estimate $z=2.6$ for $J_2 = 0.5$ and $z=2.53$ for $J_2 = 0.7$. The dashed lines in (b) show the fittings of singlet gap $\sim N^{-\frac{z}{2}}$, from which we estimate $z=1.57$ for $J_2 = 0.5$ and $z=2.16$ for $J_2 = 0.7$.}
\label{fig:gap_apdx}
\end{figure}

In Fig.~\ref{fig:gap_apdx}, we show the ED gap data for $J_2 = 0.5, 0.7$ and $\Delta = 1.0$.
The dashed lines denote the algebraic fitting of the ED data.
Since the spin triplet and singlet gaps have close amplitudes, we fit both of them. 
For triplet gap, the fitted dynamic exponents for $J_2 = 0.5$ and $0.7$ are very close, i.e. $z \simeq 2.5$.
For singlet gap, the obtained dynamic exponents are $z = 1.57$ and $2.16$ for $J_2 = 0.5$ and $0.7$, respectively.
The relatively larger difference may be owing to the stronger finite-size effect or cluster geometry reason in the calculation of singlet gap.
Overall, our analyses from the ED data suggest the dynamic exponent might be close to $2$ in the RS phase.
Such a dynamic exponent seems to be close to the obtained values in the random $J-Q$ model~\cite{liu2018}, which calls for further studies on larger system size to determine the dynamic exponent in this RS phase and whether it agrees with the result in the random $J-Q$ model.

\section{Procedure to obtain the covering of the different clusters on the lattice}

Here, we clarify that the procedure is executed on a given random sample in the middle $L_y \times L_y$ subsystem of the $L_x = 2 L_y$ cylinder. 
First of all, all the bonds including the long-range bonds are added to a list and ordered from the smallest to the largest according to their spin correlation function if the value $C_{ij}$ is smaller than a given value $C_{\rm d}$. 
In the procedure we take $C_{\rm d} = - 0.25$, because the entanglement entropy between two spins vanishes when their correlation function $C_{ij} > - 0.25$, which is impossible for them to form a ``singlet-dimer''. 
The bonds are then regarded as ``singlet-dimers'' according to their order on the list. 
There are two rules on the dimer identification process: (\romannumeral1) the bonds which have the smaller correlation functions are more likely to form singlet-dimers; (\romannumeral2) one site can only be involved in one singlet-dimer. Considering the frustration effect, there is a possibility that the singlet dimer-coverings resonance to a slightly different configuration if two bonds close to each other have nearly the same correlation function. 
Therefore, we put the bonds that could have the local resonance in different groups which we call ``resonating-dimer clusters'' and one such cluster contains $3$ spins or $2$ bonds at least.

The procedure for cluster generation is based on the above dimer process, since we still have to identify the singlet-dimer before adding it into a resonating-dimer cluster. 
A cluster is always generated from one isolated singlet-dimer but for more general explanation we assume that a cluster has already included $m$-spins with the maximum correlation function $C_{\rm max}$ of the singlet-dimer bonds involved in the cluster. 
Then, we start to traverse the entire singlet-dimer list obtained according to the above rules. 
A singlet-dimer will be added to the $m$-spins cluster so long as it involves at least one site of the cluster and satisfies the condition $C_{\rm k} - C_{\rm max} < \delta$ ($C_{\rm k}$ is the correlation function of the bond on the ordered list). 
Here, we take $\delta = 0.03125$ to be consistent with the definition in Ref.~\onlinecite{kawamura2019}.
Every time a new singlet-dimer is included , we update the maximum correlation function $C_{\rm max}$ and then repeat the process until there is no other singlet-dimer on the list can be added into the cluster.  
After the above generation procedure,  there are still some spins that are involved in neither isolated singlet-dimers nor resonating-dimer clusters.  
These unpaired spins are identified as ``orphan spins''. 
So far we have classified the three types of clusters, and each spin must be included in one of them.
Please find more details of this procedure in Ref.~\onlinecite{kawamura2019}.

\end{document}